\documentstyle[12pt]{article}
\let\useblackboard=\iftrue
\iftrue

\fi
\useblackboard
\font\blackboard=msbm10 scaled \magstep1
\font\blackboards=msbm7
\font\blackboardss=msbm5
\newfam\black
\textfont\black=\blackboard
\scriptfont\black=\blackboards
\scriptscriptfont\black=\blackboardss
\def\Bbb#1{{\fam\black\relax#1}}
\else
\def\Bbb{\bf}
\fi

\def\@cite#1#2{\if@tempswa [#1]\else$^{\scriptscriptstyle
\mbox{\rm\scriptsize#1}}$\fi}
\newcommand{\eqn}{\begin{eqnarray}}
\newcommand{\enq}{\end{eqnarray}}
\newcommand{\eqa}{\begin{array}}
\newcommand{\ena}{\end{array}}
\newcommand{\eq}{\begin{equation}}
\newcommand{\en}{\end{equation}}

\newcommand{\no}{\nonumber}

\def\comments#1{}
\newcommand{\IZ}{{\Bbb{Z}}}

\def\1N{$1\over N$}
\def\CC{\Bbb{C}}

\def\CPN{$\Bbb{C}P^{N-1}$}
\def\Gt{$\Bbb{G}(2,N)$}
\def\G24{$\Bbb{G}(2,4)$}

\def\Gkn{$\Bbb{G}(k,N)$}

\def\RR{\Bbb{R}}

\def\del{\partial}
\def\half{{1\over 2}}
\def\Tr{{\rm Tr\ }}
\def\im{{\rm Im\hskip0.1em}}
\def\bra#1{{\langle}#1|}
\def\ket#1{|#1\rangle}
\def\vev#1{\langle{#1}\rangle}

\def\Dslash{\rlap{\hskip0.2em/}D}

\catcode`\@=11
\long\def\@makefntext#1{ 
\protect\noindent \hbox to 3.2pt {\hskip-.9pt
$^{{\ninerm\@thefnmark}}$\hfil}#1\hfill} 

\def\thefootnote{\fnsymbol{footnote}}
 \def\@makefnmark{\hbox to 0pt{$^{\@thefnmark}$\hss}}  

\def\ps@myheadings{\let\@mkboth\@gobbletwo
\def\@oddhead{\hbox{} 
\rightmark\hfil\ninerm\thepage}
\def\@oddfoot{}\def\@evenhead{\ninerm\thepage\hfil 
\leftmark\hbox{}}\def\@evenfoot{}
\def\sectionmark##1{}\def\subsectionmark##1{}}

\begin{document}
\hfill{\vbox{\hbox{{\sc OUTP-94}-08P}}}
\vspace{0.5cm}
\newcommand{\symbolfootnote}{\renewcommand{\thefootnote}
	{\fnsymbol{footnote}}}
\renewcommand{\thefootnote}{\fnsymbol{footnote}}
\newcommand{\alphfootnote}
	{\setcounter{footnote}{0}
	 \renewcommand{\thefootnote}{\sevenrm\alph{footnote}}}


\newcounter{sectionc}\newcounter{subsectionc}\newcounter{subsubsectionc}
\renewcommand{\section}[1] {\vspace{0.6cm}\addtocounter{sectionc}{1}
\setcounter{subsectionc}{0}\setcounter{subsubsectionc}{0}\noindent
	{\bf\thesectionc. #1}\par\vspace{0.4cm}}
\renewcommand{\subsection}[1] {\vspace{0.6cm}\addtocounter{subsectionc}{1}
	\setcounter{subsubsectionc}{0}\noindent
	{\it\thesectionc.\thesubsectionc. #1}\par\vspace{0.4cm}}
\renewcommand{\subsubsection}[1]
{\vspace{0.6cm}\addtocounter{subsubsectionc}{1}
	\noindent {\rm\thesectionc.\thesubsectionc.\thesubsubsectionc.
	#1}\par\vspace{0.4cm}}
\newcommand{\nonumsection}[1] {\vspace{0.6cm}\noindent{\bf #1}
	\par\vspace{0.4cm}}

\newcounter{appendixc}
\newcounter{subappendixc}[appendixc]
\newcounter{subsubappendixc}[subappendixc]
\renewcommand{\thesubappendixc}{\Alph{appendixc}.\arabic{subappendixc}}
\renewcommand{\thesubsubappendixc}
	{\Alph{appendixc}.\arabic{subappendixc}.\arabic{subsubappendixc}}

\renewcommand{\appendix}[1] {\vspace{0.6cm}
        \refstepcounter{appendixc}
        \setcounter{figure}{0}
        \setcounter{table}{0}
        \setcounter{equation}{0}
        \renewcommand{\thefigure}{\Alph{appendixc}.\arabic{figure}}
        \renewcommand{\thetable}{\Alph{appendixc}.\arabic{table}}
        \renewcommand{\theappendixc}{\Alph{appendixc}}
        \renewcommand{\theequation}{\Alph{appendixc}.\arabic{equation}}
        \noindent{\bf Appendix \theappendixc #1}\par\vspace{0.4cm}}
\newcommand{\subappendix}[1] {\vspace{0.6cm}
        \refstepcounter{subappendixc}
        \noindent{\bf Appendix \thesubappendixc. #1}\par\vspace{0.4cm}}
\newcommand{\subsubappendix}[1] {\vspace{0.6cm}
        \refstepcounter{subsubappendixc}
        \noindent{\it Appendix \thesubsubappendixc. #1}
	\par\vspace{0.4cm}}

\def\abstracts#1{{
	\centering{\begin{minipage}{30pc}\tenrm\baselineskip=12pt\noindent
	\centerline{\tenrm ABSTRACT}\vspace{0.3cm}
	\parindent=0pt #1
	\end{minipage} }\par}}

\newcommand{\bibit}{\it}
\newcommand{\bibbf}{\bf}
\renewenvironment{thebibliography}[1]
	{\begin{list}{\arabic{enumi}.}
	{\usecounter{enumi}\setlength{\parsep}{0pt}
\setlength{\leftmargin 1.25cm}{\rightmargin 0pt}
	 \setlength{\itemsep}{0pt} \settowidth
	{\labelwidth}{#1.}\sloppy}}{\end{list}}

\topsep=0in\parsep=0in\itemsep=0in
\parindent=1.5pc

\newcounter{itemlistc}
\newcounter{romanlistc}
\newcounter{alphlistc}
\newcounter{arabiclistc}
\newenvironment{itemlist}
    	{\setcounter{itemlistc}{0}
	 \begin{list}{$\bullet$}
	{\usecounter{itemlistc}
	 \setlength{\parsep}{0pt}
	 \setlength{\itemsep}{0pt}}}{\end{list}}

\newenvironment{romanlist}
	{\setcounter{romanlistc}{0}
	 \begin{list}{$($\roman{romanlistc}$)$}
	{\usecounter{romanlistc}
	 \setlength{\parsep}{0pt}
	 \setlength{\itemsep}{0pt}}}{\end{list}}

\newenvironment{alphlist}
	{\setcounter{alphlistc}{0}
	 \begin{list}{$($\alph{alphlistc}$)$}
	{\usecounter{alphlistc}
	 \setlength{\parsep}{0pt}
	 \setlength{\itemsep}{0pt}}}{\end{list}}

\newenvironment{arabiclist}
	{\setcounter{arabiclistc}{0}
	 \begin{list}{\arabic{arabiclistc}}
	{\usecounter{arabiclistc}
	 \setlength{\parsep}{0pt}
	 \setlength{\itemsep}{0pt}}}{\end{list}}

\newcommand{\fcaption}[1]{
        \refstepcounter{figure}
        \setbox\@tempboxa = \hbox{\tenrm Fig.~\thefigure. #1}
        \ifdim \wd\@tempboxa > 6in
           {\begin{center}
        \parbox{6in}{\tenrm\baselineskip=12pt Fig.~\thefigure. #1 }
            \end{center}}
        \else
             {\begin{center}
             {\tenrm Fig.~\thefigure. #1}
              \end{center}}
        \fi}

\newcommand{\tcaption}[1]{
        \refstepcounter{table}
        \setbox\@tempboxa = \hbox{\tenrm Table~\thetable. #1}
        \ifdim \wd\@tempboxa > 6in
           {\begin{center}
        \parbox{6in}{\tenrm\baselineskip=12pt Table~\thetable. #1 }
            \end{center}}
        \else
             {\begin{center}
             {\tenrm Table~\thetable. #1}
              \end{center}}
        \fi}


\def\@citex[#1]#2{\if@filesw\immediate\write\@auxout
	{\string\citation{#2}}\fi
\def\@citea{}\@cite{\@for\@citeb:=#2\do
	{\@citea\def\@citea{,}\@ifundefined
	{b@\@citeb}{{\bf ?}\@warning
	{Citation `\@citeb' on page \thepage \space undefined}}
	{\csname b@\@citeb\endcsname}}}{#1}}

\newif\if@cghi
\def\cite{\@cghitrue\@ifnextchar [{\@tempswatrue
	\@citex}{\@tempswafalse\@citex[]}}
\def\citelow{\@cghifalse\@ifnextchar [{\@tempswatrue
	\@citex}{\@tempswafalse\@citex[]}}
\def\@cite#1#2{{$\null^{#1}$\if@tempswa\typeout
	{IJCGA warning: optional citation argument
	ignored: `#2'} \fi}}
\newcommand{\citeup}{\cite}

\def\@cite#1#2{\if@tempswa [#1]\else$^{\scriptscriptstyle
\mbox{\rm\scriptsize#1}}$\fi}

\def\fnm#1{$^{\mbox{\scriptsize #1}}$}
\def\fnt#1#2{\footnotetext{\kern-.3em
	{$^{\mbox{\sevenrm #1}}$}{#2}}}

\font\twelvebf=cmbx10 scaled\magstep 1
\font\twelverm=cmr10 scaled\magstep 1
\font\twelveit=cmti10 scaled\magstep 1
\font\elevenbfit=cmbxti10 scaled\magstephalf
\font\elevenbf=cmbx10 scaled\magstephalf
\font\elevenrm=cmr10 scaled\magstephalf
\font\elevenit=cmti10 scaled\magstephalf
\font\bfit=cmbxti10
\font\tenbf=cmbx10
\font\tenrm=cmr10
\font\tenit=cmti10
\font\ninebf=cmbx9
\font\ninerm=cmr9
\font\nineit=cmti9
\font\eightbf=cmbx8
\font\eightrm=cmr8
\font\eightit=cmti8


\centerline{\tenbf SOLVABLE MODELS IN TWO-DIMENSIONAL N=2 THEORIES}
\baselineskip=22pt
\vspace{0.8cm}
\centerline{\tenrm Mich\`ele Bourdeau}
\baselineskip=13pt
\centerline{\tenit Dept. of Theoretical Physics, Oxford University}
\baselineskip=12pt
\centerline{\tenit Oxford, OX1 3NP, U.K.}
\vspace{0.3cm}
\vspace{0.9cm}
\abstracts{N=2 supersymmetric field theories in two dimensions have been
extensively studied in the last few years. Many of their properties
can be determined along the whole renormalization group flow, like
their coupling dependence and soliton spectra. We discuss here several
models  which can be solved
completely, when the number of superfields is taken to be large,
by studying their
topological-antitopological fusion equations. These models are the
\CPN model, $\sigma$-models on Grassmannian manifolds, and certain
perturbed $N=2$ Minimal model.}

\vfil
\twelverm\baselineskip=14pt
\section{Introduction}

Two-dimensional quantum field theories have been studied extensively
in the last few years, both at the conformal point and off criticality
for massive and massless flows.
More recently, there has been an extensive exploration of
two-dimensional field theories with $N=2$ supersymmetry. These
theories play for example a fundamental role in the
construction of $N=1$ string
vacua and the study of string compactification.
Many of the models admit Landau-Ginsburg-type actions characterized by
a superpotential which obeys non-renormalization theorems, thus
making them easier to study. The superpotential encodes the chiral
ring of operators (the topological data of the theory) and many of the
properties of the full quantum theory can be determined through it by
defining a metric, the inner product on the space of supersymmetric
ground-states (a generalization of Zamolodchikov's metric) which
satisfies a set of differential equations on the space of couplings,
the topological-antitopological ($tt^*$) fusion equations. This
metric, and a new index related to it, are defined along the whole
renormalization
group flow and give information about
the  soliton spectrum and other characteristics of the model such as
the scale and coupling dependence.
The new index, a non-perturbative quantity, is calculable exactly in
any two-dimensional $N=2$ theory, whether or not it is integrable.
{}From it  a sort of $c$ function on
the space of couplings can also be defined, which decreases
monotonically along the flow.
 The `new index' can be calculated either from the ground-state metric
through the $tt^*$ equations, or by solving the TBA equations if the
S-matrix of the model is known.
The $tt^*$ fusion equations have proven
difficult to solve and their equivalence with the TBA approach has
only been showed numerically for some simple cases.

The $tt^*$ formalism has a wide range of applications in
two-dimensional systems. Apart from the models we will be discussing
here,
many of the properties of two-dimensional polymers (self-avoiding random
walks) can be studied. The index turns out to be the partition
function for a single polymer which loops around a cylinder and it gives
the scaling function for the number of such configurations.

Here, we  investigate several models where the $tt^*$ equations
can be solved completely analytically.
Quantum field theories usually become tractable  when
the number of fields in the model is very large. For example, the
large $N$ \CPN model has been  solved: to leading order
in $1/N$, $S$-matrix elements are given by summing tree
diagrams, while bulk quantities like the free energy are calculable by
simply extremizing an effective action.

We  show here that the $tt^*$ methods show comparable
simplifications in this limit. The models we will be considering
 are the  \CPN\ model,
 $\sigma$-models on Grassmannian target manifolds \Gkn\ and
 certain perturbed $A_n$ Minimal models.


\section{The ground-state metric for N=2 theories and the new index}

The situation governed by the $tt^*$ equations is the following.
We have a $d=2$, $N=2$ supersymmetric field theory quantized on a
Euclidean manifold, with metric and boundary conditions preserving $N=2$
global
supersymmetry.
We assume there are a discrete set of supersymmetric ground states, that
at
least one dimension is compact, and that a Hamiltonian defined on a
compact
hypersurface has a gap.
Given all this, certain  correlation functions can be
reduced to
sums over the ground states.

One way to describe the $tt^*$ results
is
that they answer the question: from the topological data of an $N=2$
theory, in which only the ground-states survive,
what can one reconstruct about the original theory?

Let us then consider an $N=2$ theory on a cylinder, where the long
dimension
has  length $L$ (which will go to
infinity), and the compact dimension has
circumference
$\beta$.

In two dimensions the $N=2$ supersymmetry algebra
takes the form
\eqn\label{algebra}
Q^{+2}_L=Q^{-2}_L={Q}^{+2}_R&={Q}^{-2}_R=&\{Q^+_L,{Q}^-_R\}=
\{Q^-_L,{Q}^+_R\}=0\no\\
\{Q^+_L,{Q}^+_R\}=2\Delta&  & \{Q^-_L,{Q}^-_R\}=2\Delta^*\no \\
\{Q^+_L,Q^-_L\}=H-P &  & \{Q^+_R,Q^-_R\}=H+P \no\\
{}[F, Q^{\pm}_L]=\pm Q^{\pm}_L & &  [F,{Q}^{\pm}_R]=\mp{Q}^{\pm}_R \no\\
Q^{+\dagger}_L=Q^-_L & &  ({Q}^+_R)^\dagger = {Q_R}^-\\
Q^{\pm}\equiv{1\over\sqrt{2}}(Q^{\pm}_L+{Q}^{\pm}_R)& &\{Q^-,Q^+\} = H.
\enq
$F$, the `fermion number,' is the conserved $U(1)$ charge, and
$Q^{\pm}_L$  and ${Q}^{\pm}_R$ are left and right supercharges.
In non-compact space one can have a non-zero central term
$\Delta$, which will come in below.

The most basic elements particular to a given $N=2$ theory are the chiral
and anti-chiral rings.
The chiral operators
$\phi_i$ satisfy $[Q^+,\phi_i]=0$,
and the anti-chiral operators $\bar{\phi_i}$ satisfy
$[Q^-,\bar{\phi_i}]=0$.
The chiral ring is defined in terms of the operator product algebra as
\eq
    \phi_i\phi_j  =  \sum_k C^k_{ij}\phi_k + [Q^+,\Lambda].
\en
Since the derivative of any operator (and the stress tensor itself) is a
descendant under $Q^+$ (and $Q^-$), the positions of the operators on the
left
hand side do not matter.
The anti-chiral ring will have structure constants $\bar C^k_{ij}$.

Equally important are the supersymmetric (Ramond) ground states
\eq\label{susygr}
H|a\rangle = Q^{\pm}|a\rangle = 0.
\en
We could make a correspondence between these and chiral fields by choosing
a
canonical ground state $\ket{0}$.
Then we can identify
\eq
               \phi_i|0\rangle=|i\rangle+Q^{+}|\Lambda\rangle .
\en
Finally we could project on the true ground state by applying an operator
like
$\lim_{T\rightarrow\infty} \exp -HT$.
We could also do this with anti-chiral fields
$\bar{\phi}_i$, producing states to be called $|\bar{i}\rangle$.
The structure constants $C^k_{ij}$ then also give the action of the chiral
operators on the ground states:
\eq\label{chiralact}
  \phi_i|j\rangle  =  C^k_{ij}|k\rangle + Q^+|\psi\rangle.
\en

This construction is not completely satisfactory because it is not clear
that
the correspondence is one to one; furthermore it depended on the choice of
$\ket{0}$.
Both problems are dealt with by making a correspondence using
spectral flow.\cite{cv1}
In principle, this constructs the state $\ket{i}$ by doing a
path integral on a hemisphere with an insertion of $\phi_i$.
We need spectral flow to put this state in the Ramond sector, and we can
think
of it as turning on a $U(1)$ gauge field coupled to the fermion number
current,
with holonomy $e^{i\pi}$ on the boundary.
We can then take as $\ket{0}$ the state produced by inserting the identity
operator $\phi_0\equiv 1$, and non-degeneracy of the two-point function
$\vev{\bar\phi_i \phi_j}$ will imply that the correspondence is one to
one.

Now we take $|i\rangle$ and $|\bar{j}\rangle$ to denote the
basis of
ground states corresponding to the fields $\phi_i$ and $\phi_{\bar{j}}$.
CPT will relate $|i\rangle$ to a state $\bra{\bar i}$ so the usual Hilbert
space metric will be the hermitian
\eq\label{metricg}
g_{i\bar{j}}=\langle\bar{j}|i\rangle.
\en
Another structure present in the theory is the ``real structure'' $M$
expressing one basis in terms of the other:
\eq\label{metricM}
    \langle\bar{i}|=\langle j|M^j_{\bar{i}}
\en
CPT implies $MM^*=1$.

\noindent A combination of these produces the `topological' metric $\eta$:
$ \quad    g_{i\bar{k}}=\eta_{ij}M^j_{\bar{k}}.\quad $
This is the two-point function in the topologically twisted theory.

Supersymmetry-preserving perturbations of the action are of two types.
In general we need to write a commutator with all four supercharges
(or integral $d^4\theta$) to preserve all supersymmetries.
However we can also write
\eq\label{pertS}
     \delta S =\sum_i
\int d^2x~~ \delta t_i\{Q^-_R,[Q^-_L,\phi_i]\}
+\delta \bar{t}_i \{Q^+_R,[Q^+_L,\bar\phi_i]\}.
\en
where $\phi_i$ and $\bar\phi_i$ are chiral and anti-chiral fields.
A perturbation which can only be written in this form is called an
$F$ term; the others are $D$ terms.

In the spirit of the non-abelian Berry's phase, define the gauge
connection
\eq
   A^{\hphantom{i}j}_{i\hphantom{j}k}=
g^{j\bar j'}\langle\bar{j'}|\del_i|k\rangle
\en
and its conjugate.
By definition, the metric $g$ is covariantly constant with respect to the
derivatives
$\quad D_i=\del_i-A_i,\quad \bar{D}_i=\bar{\del}_{\bar i}-\bar{A}_{\bar i}.$

\noindent Then, we might expect covariant combinations like the
curvature to be especially simple.
Writing these out explicitly and manipulating the supercharges gives
\eq\label{holo}
 [D_i,D_j]  =  [ \bar{D}_{\bar i},\bar{D}_{\bar j}]=0
\en
and for the mixed terms, one finds
\eq
 [D_i,\bar{D}_{\bar j}]  = -\beta^2[C_i,\bar{C}_{\bar j}],
\en
a differential equation for the metric.
By (\ref{holo}) one can choose a basis in which
$\bar{A}_i=0$, so $A_i=g^{-1}\partial_ig$, and (14) becomes
\eq
  \bar{\partial}_{\bar j}(g\partial_ig^{-1})
  =\beta^2[C_i,gC^{\dagger}_{\bar j}g^{-1}]
\en
These are the $tt^*$ equations, which given enough boundary conditions
determine the metric $g$.
Different models with the same chiral ring can have different metrics:
thus the boundary conditions are a crucial part of the story.
In the cases considered in detail,\cite{cv1,cv3,cfiv}
the dependence on one relevant coupling is studied.  Let this define a
mass
scale $m$; then small $\beta m$ is weak coupling and this limit of the
metric
can be found using semiclassical techniques.
The large $\beta m$ boundary conditions are even simpler and are best
explained
in terms of the `new index' (see below).

There is another observable depending only on $F$ couplings\cite{cfiv}.
Although it is simply related to the metric it has a
clearer
physical interpretation.
It is modeled after the index ${\rm Tr} (-1)^F e^{-\beta H}$,
which is completely
independent of finite perturbations of the theory for $N\geq 1$
supersymmetric
theories in any dimension.\cite{wit3}
This index has been very useful in providing criteria for
supersymmetry breaking.
For an $N=2$ theory in two dimensions,
the new  observable is the `new index' $\,\,{\rm Tr}\, F(-1)^Fe^{-\beta
H},\,\,$ with $F$ the `Fermion number.'

The new index is actually a matrix since the boundary conditions at
spatial
infinity can be any vacuum of the theory. Let the left vacuum be $a$ and
the
right one $b$, and consider the matrix elements
\eq\label{newindex}
    Q_{ab} ={{i\beta}\over{L}}{\Tr}_{ab}\;(-1)^F\; F\;e^{-\beta H}.
\en
In \cite[x]{cfiv} it is shown that the matrix $Q$ is imaginary and
hermitian,
and that
\eq\label{metrel}
      Q_{ab} 
             = i(\beta g{\del_\beta}g^{-1} + n)_{ab}
\en
where $n$ is the coefficient of the chiral anomaly.

This quantity is
particularly suited for extracting the soliton spectrum and other low
temperature properties of the model.
The simplest case is a model with a mass gap; clearly $Q$ will be
exponentially
small in $\beta m$ and typically each of the leading terms in an expansion
in
$\exp -\beta m$ is the contribution of a single massive
particle saturating the Bogomolnyi bound $m=|\Delta|$.
\cite{fend,cfiv,witolive}

\section{\CPN models}

\noindent We now apply the formalism to the supersymmetric \CPN
$\sigma$-models.

\noindent Non-linear $\sigma$-models define maps
from spacetime into a riemannian
target manifold $M$.  Supersymmetric $d=2$ $\sigma$-models exist for any
target
manifold.  If the target manifold and metric is K\"ahler,\cite{zu}
the model will be $N=2$.

\noindent A manifestly $N=2$ invariant superspace Lagrangian
is
\eq
{\cal L}=\half\int dx\;d^2\theta \;d^2\bar{\theta}\;\;
K(\Phi,\Phi^{\dagger}),
\en
where $K$ is the K\"ahler potential
and $\Phi_i$ are complex chiral superfields
\eq
\Phi_i=\Phi_i(x,\theta,\bar{\theta})=\varphi_i(x)+\sqrt{2}
\epsilon_{\alpha\beta}\theta^{\alpha}\Psi_i^{\beta}(x)+
\epsilon_{\alpha\beta}\theta^{\alpha}\theta^{\beta}F_i(x)~.
\en
Any term in $K$ which is globally defined on $M$ is a $D$ term, and
conversely
two choices of $K$ for which the K\"ahler forms $J=dz^i \wedge d\bar
z^{\bar
j}\del_i{\bar{\del_{\bar j}}} K$ are in different complex cohomology
classes
differ by $F$ terms.
For \CPN, $\dim H^{1,1}(M,\RR) = 1$ and the K\"ahler class is specified by
a
single parameter.  We can take
\eq\label{kahler}
K(\Phi,\Phi^{\dagger})={1\over{g^2}}
\log(1+\sum_{i=1}^{N-1}\Phi_i^{\dagger}\Phi_i).
\en

The supersymmetric ground states of an $N=2$ $\sigma$-model are in one-to-one
correspondance with the complex cohomology classes of the target space,
and by spectral flow so are the chiral primaries.
Using semiclassical techniques to compute the chiral ring, one finds it to
be a deformation of the classical cohomology ring:
instantons can contribute to correlation functions of the chiral
primaries.
In simple cases the possible contributions are determined by the
chiral anomaly.
For \CPN, the classical cohomology ring is the powers of the K\"ahler form
$x$ allowed on a $2(N-1)$-dimensional manifold, up to $x^{N-1}$.
The instanton changes the relation $x^{N}=0$ to
\eqn\label{chiral}
x^{N}=e^{-2\pi/g^2},
\enq
which defines the chiral ring.

Many $N=2$ supersymmetric theories in two dimensions admit a
Landau-Ginsburg description\cite{lvw,warner}
 if their superspace Lagrangian
is of the form
\eq\label{lglang}
{\cal L}=\int d^4\theta \sum_i \phi_i\bar{\phi}_i +\int d^2\theta
W(\phi_i)+h.c.
\en
where $\phi_i$, $\bar{\phi_i}$ are the chiral and antichiral
$(a,c)$ superfields and the superpotential $W$  is an
analytic function of the complex superfields which obeys non-renormalization
theorems.
The ground states of the theory are  $dW(\phi)=0$.
The chiral ring is the ring of polynomials generated by the
$\phi_i$
modulo the relations
$dW(\phi)/d\phi_i = D \bar D \phi_i \sim 0$.

In this light, and for many other purposes, a more useful definition
of the \CPN $\sigma$-model is provided
by a gauged $N=2$ model, which constructs \CPN as a quotient of $\CC^N$:
\eq\label{lgform}
   {\cal L} = \int d^4\theta \left [ \sum_{i=1}^N \bar{S}_i
    e^{-V} S_i + {N\over{g^2}} V\right ] .
\en
$S_i$ are $N$ chiral superfields which become
the homogeneous coordinates on \CPN.
We have introduced a factor of $N$ with the coupling $1/g^2$ which will
make
the $N\rightarrow\infty$ limit well defined.\cite{coleman}
$V$ is a real vector superfield, whose components become the many
auxiliary
fields of the following component form of the Lagrangian:
\eqn\label{compform}
{\cal L}&=&{N\over g^2}\bigl\{\ (D_\mu n^*_i)^\dagger
  (D_\mu n_i)+\bar{\psi}^i(i\Dslash+\sigma+i\pi\gamma^5)\psi_i- \no \\
        &&+(\sigma^2+\pi^2)-\lambda(n^*_in_i-1)
	+ \bar\chi n^*_i \psi^i + \bar\psi^i n_i\chi \ \bigr\}.
\enq
The superfields $S$ have complex components $n_i$ and $\psi_i$.
(We rescaled them by $\sqrt{N}/g$.)
The constraint $n^*_in_i=1$ is imposed by the Lagrange multiplier
$\lambda$;
the phase of $n$ and $\psi$ is gauged by $A_\mu$ (which appears in
$D_\mu=\del_\mu+iA_\mu$).
The fields $\sigma$ and $\pi$ implement 4-fermi interactions and by the
equations of motion are equal to $\bar\psi^i\psi_i$ and
${i}\bar\psi^i\gamma^5\psi_i$ respectively.

 Integrating out the superfields $S_i$ in (\ref{lgform}),
one obtains an effective action for the \CPN models  which has the
form of a Landau-Ginsburg model. By gauge invariance
only the field-strength superfields $X$ and $\bar{X}$ remain since $V$ is
not gauge invariant:
\eqn
S_{\rm eff}={N\over{2\pi}}\!\int\!\!d^2x\left\{
\int \!\! d^2\theta W(X)\!+\!\!
\int\!\! d^2\theta \bar{W}(\bar{X})
\!+\!\!\int\!\! d^4\theta[Z(X,\bar{X},\Delta,\bar{\Delta})]\right\}
\enq
with
\eq
W(X)= X(\log X^N-N+A(\mu)-i\theta).
\en
$A$ is a renormalized coupling and $\theta$  the
instanton angle,
and
\eq
  X=D_L\bar{D}_R V,\qquad\bar{X}=D_R\bar{D}_L V.
\en
The chiral ring is the powers of $X$ mod $dW=0$:\quad
$X^N = e^{ -A+i\theta}.$

\subsection{Topological-antitopological fusion equations and the new index}
\vspace*{-0.35cm}
The action can be written in the following form
\eq
S=-{N\over 4\pi}\ln t\int d^2yd^2\theta \; x + {\rm c.c.}
\en
where  $x=x(\Phi_i,\bar{\Phi}_i)=
\bar{D}D\ln(1+\sum\Phi_i\bar{\Phi}_i)$ represents
the K\"ahler class, $\ln t~ x$ is the K\"ahler form and
$\Phi_i,\bar{\Phi}_i$ are chiral superfields.

The chiral ring is generated
by a single element $X$, the K\"ahler class, with relation
\eq
X^N=t^N
\en
\noindent To write down the $tt^*$ equations,
we need to find the operator corresponding to a perturbation of $t$,
and its action on the chiral ring $(X^{N-1},\ldots,X,1)$.
It is represented by the matrix
\[ C_t
={N\over4\pi t}\left(\eqa{cccccc}
0 & 1 & 0 & \dots & 0 & 0 \\
0 & 0 & 1 & \dots & 0 & 0 \\
\dots & \dots & \dots & \dots & \dots & \dots \\
\dots & \dots & \dots & \dots & \dots & \dots \\
0 & 0 & 0 & \dots & 0 & 1 \\
t^N & 0 & 0 & \dots & 0 & 0
\ena \right) \]
\noindent The $\IZ_N$ symmetry implies that the metric
$g_{i\bar{j}}=\langle\bar j|i\rangle$ is diagonal.
The metric $g$ is a function only of $|t|^2$, because it is a path
integral
with total
chiral charge zero, and chiral charge non-conservation is proportional to
instanton number.
Thus the equations become o.d.e.'s in terms of $|t|$.
We can write them in terms of the dimensionless parameter
$x=\beta t/2\pi$, but to do this we need to take out the dimensional
factors
in $g_{i\bar j}$.
Thus we define
\eq
         q_j  =  \ln g_{j\bar{j}} + 2j\log|t| \qquad\qquad
          q_{j+N} \equiv q_j.
\en
We will now use $x$ as our coupling, and call it $\beta$
in the following.  The $tt^*$ equation  becomes
\eq\label{ttstar}
    {4\over N^2} \partial_\beta\partial_{\beta^*}q_i +
          e^{(q_{i+1}-q_i)} -
        e^{(q_i-q_{i-1})}=0.
\en
This equation is the affine $\hat{A}_{N-1}$ Toda equation.

A solution should be determined by
the boundary conditions near $\beta\sim 0$ and $\beta\sim \infty$.
In \cite[x]{cv1} these are found explicitly for the small $\beta$ limit
by a
semiclassical calculation of the metric.  For the large $\beta$ limit
the solution is exponentially
small
in $\beta$, the precise form of the leading exponential being
determined
by the soliton spectrum, which is already known for these (integrable)
models.
For the cases of $\CC P^1$ and $\CC P^2$, the $tt^*$ equations become
special cases of the Painlev\'e ${\rm I\!I\!I}$ equation, and the
connection formula between small and large $\beta$
asymptotics is known.\cite{cv2}.

A reasonable ansatz for the large $N$ limit would be that the metric and
index
are continuous functions of the variable $s\equiv i/N$.
The large $N$ effective action of the \CPN\ model still has $N$ fields
with an action of the form exp($NS_{eff}$) at an appropriate saddle
point. We therefore rescale  $q_i \rightarrow {1\over N}~q_i$.
The $tt^*$ equation becomes
\eq\label{oureqn1}
       4\partial_{\beta}\partial_{\beta^*}H +{{\partial^2}\over
      {\partial s^2}} e^H = 0,
\en
with $H=q'$.

This equation has been studied in several contexts.
It was first noted for a connection with 4D self-dual
gravity.\cite{bf,gd,eh}
More recently, it has been
studied in the context of the large $n$ limit of $W_n$
algebra.\cite{bakas,bg,park}
It is also a well known scaling limit of the two-dimensional
infinite Toda lattice.\cite{saveliev}
A formal solution of the boundary value
(Gours\'at) problem for the equation has been given
in \cite[x,y]{kssv}.

We still need to specify the boundary conditions to select a solution to
the equation.
We can deduce the large $N$ limit of our metric for small $|\beta|$
from the semi-classical result of Cecotti and Vafa\cite{cv3}
(this is essentially the two-point function $\vev{\phi_i\bar\phi_j}$
reduced to constant field configurations, or
$\int d\varphi~x^i \wedge *x^j$)
\eq
\label{smallbeta}
e^H=
{s(1-s)\over|\beta|^2(-\ln (|\beta|/2) - \gamma)^{2}}
\en
(for $0<s<1$ and defined elsewhere by periodicity), where $\gamma$
is Euler's constant, a factor predicted by the connection
formula for the $n=1,2$ equations, and described in \cite[x]{cv3} as
a one-loop correction to the semiclassical calculation.

For large $\beta$, each sector with one soliton of mass $m $ satisfying
the
Bogomolnyi  bound contributes $((f+1)-f)\exp -\beta m$ times a factor
depending
on its central charge $\Delta$ to the new index,
and by (\ref{metrel}) to $H$.
We have $N$ such solitons with mass $m=m_0$; the
central charge is $\pi(+L)-\pi(-L)$ or one can just linearize
(\ref{oureqn1}) to
see the appropriate boundary condition
\eq
\label{largebeta}
 H(s)\sim -{{\exp{(-2\pi|\beta|)}}\over{\sqrt{2\pi|\beta|}}}
\cos (2\pi s)
\en
which satisfies our equation to  first order.
This limit is not a solution and one could use (\ref{oureqn1}) to generate
corrections to $H$ coming from multi-soliton sectors.

The new index is
\eq\label{index}
    Q_{ab} ={{i\beta}\over{L}}{\Tr}_{ab}\;(-1)^F\; F\;e^{-\beta H}
\en
where $a$ and $b$ characterize the vacua at spatial infinity.
We can rewrite it as a path integral with an insertion of
the fermion number charge $\int dx^1 J^0_F$.

Recall now the action (\ref{lgform}) and its component form (\ref{compform}).
Since the fields $S_i$ appear quadratically in (\ref{lgform}),
it is possible to integrate
them out exactly, at least in terms of a one-loop determinant:
\eqn\label{effact1}
{\cal L}&=&{-N}\Tr\log(-D_\mu D^\mu + \lambda)
+ {N}\Tr\log(i\Dslash+\sigma+i\pi\gamma^5) \no
\\
        &&+{N\over g^2}(\sigma^2+\pi^2-\lambda)
	+ {\rm fermionic}.
\enq
In the large $N$ limit,
the $N$ in front
of
the action means that the  integration over auxiliary fields can
be
done by saddle point, and calculations at leading order in $1/N$ can be
done by
classical techniques.

The effective action (\ref{effact1}) has been
extensively studied at $T=0$ \cite{dadda} and at finite temperature
(typically not in the supersymmetric context, but the results can be
easily
adapted)\cite{davis,affleck}.
In the large $N$ limit it is $O(N)$ and we can calculate bulk quantities
like
the free energy simply by extremizing it with respect to the auxiliary
fields.
The `new index' is computed similarly, with the main differences being
that
we take periodic fermi boundary conditions (and have unbroken
supersymmetry),
we insert the fermion number operator $F$ (this will be done by
differentiating
with respect to a coupling at the end), and we fix the boundary conditions
at
$x^1=\pm L$ to go to two possibly different values of
$\sigma+i\pi$ (the ground states being characterized by the phase of
$\sigma+i\pi$).\cite{bd}
This last condition means that we need to consider non-constant background
fields in the functional integral.
For general background fields this is quite complicated, but what saves us
is
that the required variation is small, of $O(1/N)$, so we only need
the
leading terms in an expansion in derivatives and amplitude.
The derivative terms (to the accuracy we need them) are
\def\loweff{
S_{eff}&=& N \int d^2x
{1\over 8\pi \mass^2} \left(F_{\mu\nu}^2 + (\partial_\mu \sigma)^2
+ (\partial_\mu \pi)^2\right)
+ {i\over 2\pi} \epsilon^{\mu\nu} F_{\mu\nu} \im\log (\sigma+i\pi)
+ V_{eff} + \ldots
}
\def\mass{m_0}
\eqn\label{eqloweff}
\loweff
\enq
The finite temperature effective potential is
\eqn
V_{eff}&=&{1\over {g^2}}(\sigma^2+\pi^2)-
\sum_{k^0} \int{{dk^1}\over{(2\pi)^2}}{\rm
tr}
\ln [k_\mu\gamma^\mu-A_\mu\gamma^\mu-(\sigma
+i\pi\gamma^5)]\no \\
&&-{1\over{g^2}}\lambda +\sum_{k^0} \int {{dk^1}\over{(2\pi)^2}}
\ln [(k_\mu-A_\mu)^2+\lambda] \label{kzerosum}\\
&=& V_{eff}|_{T=0}-\int^{\infty}_0{{dk}\over{2\pi}}
\ln|1-e^{-\beta\sqrt{k^2+\lambda}}
e^{i\beta A_0}|^2 \no\\
&&\qquad\qquad +\int^{\infty}_0{{dk}\over{2\pi}}
\ln|1-e^{-\beta\sqrt{k^2+\sigma^2}}
e^{i\beta A_0}|^2\label{fermterm1}.
\enq
This is essentially the standard expression from statistical mechanics of
a
free field \cite{kapusta} (with chemical potential $iA_0$)
with one difference: we incorporated the periodic fermion boundary
conditions,
which led to the sign change in (\ref{fermterm1}).

We will not go into the details to obtain the new index, but just
state here the results.\cite{bd}
The index is obtained  by the saddle point
approximation

\eq\label{indexr}
Q(s,\beta) = -{N\over 2\pi i}{d\over ds} {\beta\over
NL}S_F({\beta,s})|_{min}
\en
with
\eqn
	{\beta\over NL} S_F &=& -{1\over 2\pi}(2\pi (s-\half)+ \beta
A_0)^2
 + \beta^2\sigma^2 {1\over{4\pi}}\left(\ln{\sigma^2\over m_0^2}-1\right) -
\\
    &&~\beta\int^{\infty}_0{{dk}\over{2\pi}}
{}~~\ln \left|1-e^{-\beta\sqrt{k^2+\sigma^2}+i\beta A_0}
\right|^2.\no
\enq

We still need to minimize the effective action with respect to
the auxiliary fields $\sigma$ and $A_0$.
It turns out that the minimization can be reinterpreted as precisely
a combination of known techniques for finding solutions to self-dual
gravity from those for simpler equations via Legendre transform. In
this way we can prove that the index and metric, computed
independently, solve the $tt^*$ equation. We refer the interested
reader to \cite[x]{bd}.

\section{The Grassmannian models}

Now we discuss non-linear $\sigma$-models on  complex
Grassmannian target manifolds \Gkn. These spaces have
(complex) dimension
$k(N-k)$ and consist of all $k$ dimensional subspaces of the complex
vector space ${\bf C}^N$.

The classical cohomology\cite{bott,ken,gep,wit1} of the
Grassmannian manifold \Gkn\
is generated from  the Chern classes
$X_i$,  where $X_i$ is a
$(i,i)$ form, with the relations
\eq\label{coh}
\sum_{i\geq 0}X_it^i~.~\sum_{j\geq 0}Y_jt^j=1
\en
where $Y_j's$ are some functions.

 The quantum cohomology ring results from a
modification to relations
(\ref{coh}) of the form\cite{wit1}
\eq\label{ring}
\bigl(\sum_{i=0}^kX_i~t^i\bigr)~.~\bigl(\sum_{j=0}^{N-k}Y_j~t^j\bigr)
=1+(-1)^{N-k}~t^N
\en
which imply
$Y_{N+1-i}+(-1)^{N-k}~\delta_{i,1}=0,\quad 1\leq i\leq k$.

The quantum cohomology ring is generated by
polynomials in the $X_i$'s (if one eliminates the $Y_j$'s)
subject to the constraints (\ref{ring}), and has dimension
$N!/k!(N-k)!$.

As quantum field theories,   the
Grassmannian $\sigma$-models (for $k\geq 2$) can be thought of
as generalizations of the \CPN
models. The Lagrangian has a similar form
\eq\label{lgg}
   {\cal L} = \int d^4\theta \left [ \sum_a \bar{S}_a
    e^{-V} S_a + \alpha \Tr V\right ]
\en
where now the
chiral fields $S_{ia}$ carry two indices, a `gauge' $U(k)$ index $i$,
and a `flavour' $SU(N)$ index $a$,  since  there are
$(N\times k)$-matrix scalar fields $n=(n_i^a)$ and
$(N\times k)$-matrix Dirac spinor fields $\psi=(\psi^a_i)$.
$V$ is  a $k\times k$ matrix of superfields with gauge group $U(k)$
 (see \cite[x]{abd}).

Integrating out the superfields, one recovers again
 a Landau-Ginsburg-type action:\cite{cv2} The field-strength
superfields (which we now call $\lambda$, $\bar{\lambda}$)
belong to the adjoint representation of $U(k)$ and
are now gauge covariant.
The real gauge-invariant objects of (\ref{lgg})
are  the Ad-invariant polynomials in the
field-strengths $\lambda$, with a ring generated
by the $(a,c)$ superfields
$X_i\quad(i=1,2,\dots,k)$ defined by
\eq
\det [t-\lambda]=t^k+\sum_{j=1}^k(-1)^jt^{k-j}X_j.
\en
At the topological field theory level one can assume that
 the $\lambda$'s and $\bar{\lambda}$'s are
independent of each other and therefore, as matrices,  are
diagonalizable and all their eigenvalues $\lambda_m$ are distinct.
Then, without loss of generality, the
functional determinants in the path integral
are the same as for the \CPN case, since the
full background is now abelian,
 and the superpotential, which fixes the theory completely, becomes
\eq\label{lgw}
 W_f(\lambda_1,\lambda_2,\dots,\lambda_k)={1\over {2\pi}}
\sum_{j=1}^k\lambda_j(\log \lambda_j^N-N+A(\mu) -i\theta ).\no\\
\en
The gauge-invariant fields are now polynomials in the eigenvalues
$\lambda_m$ of the field-strengths $\lambda$ and are generated by the
elementary symmetric functions
\eq\label{var}
X_i(\lambda )\equiv\sum_{1\leq l_1
< l_2 < \dots l_i\leq k}\lambda_{l_1}\lambda_{l_2}
\dots \lambda_{l_i}\quad (i=1,\dots, k)
\en

The ring relations are
$\lambda_j^N=$ const. and the quantum cohomology  of the
Grassmannian $\sigma$-models will be generated by
the elementary symmetric functions $X_i$'s. As quantum field theories, the
Grassmannian $\sigma$-models \Gkn\ can thus be identified as the
 tensor product of $k$ copies of the \CPN $\sigma$-model
with certain redundant states eliminated.\cite{cv2}

We are interested in finding a suitable basis for the metric of
the Grassmannian $\sigma$-model.
In view of (\ref{lgw},\ref{var}), a basis for the chiral ring will consist of
polynomials $P_r(X_i)$ (for $ r=1,\dots , N!/k!(N-k)!$) in the $X_i$'s.
Then one can determine $C$ in (15) from the ring relations
and derive the
$tt^*$ equations. However the equations are difficult
to handle in this form. A more enlightening way to obtain and study
the equations is to use the map (\ref{var}).
The metric is then  defined in terms of the variables $\lambda_m$,
in the following way
\eq\label{ch}
\vev{P_r(X_i)|P_s(X_j)}={1\over{k!}}\vev{\Delta(\lambda)
P_r(X_i(\lambda))|\Delta(\lambda)P_s(X_j(\lambda))}_f
\en
where $\Delta(\lambda)$ is the Vandermonde determinant
and is the Jacobian $J=\det (\del X_i/\del\lambda_j)$ of the transformation
in (\ref{var}).

Each  basis element $P_s(X_j)$ can then be written as a polynomial
$\Delta(\lambda)P_s(X_j(\lambda))$ in
the different $\lambda_m$'s. The metric on the LHS  of (\ref{ch})
is  a sum over the products of the metrics
for the \CPN  models, which are
diagonal $\vev{\overline{\lambda_m^k}|\lambda_m^s}=\delta_{ks}
\vev{\overline{\lambda_m^k}|\lambda_m^s}$.

The ground-state metric is in general non-diagonal and complicated for
an arbitrary basis. However,
the form of the metric  suggests
that one might try finding an orthogonal basis
in which the metric becomes diagonal and simple (i.e. with each
component consisting of a single term).
This basis is the `flat coordinates' basis, in which
the $tt^*$ equations have the simplest form\cite{cv1}. This basis
is characterized by a  two-point function $\eta$ which
is independent of the perturbing
parameters of the model and  squares to 1  ($\eta^*=\eta^{-1}=\eta$).

For example, for \Gt,  such a basis is given by
the   $N(N-1)/2$ elements\\
$\ket{mn}\equiv\ket{\lambda^m\lambda^n}$ with $n>m$,  such that
$\,\,\ket{mn}=-\ket{nm}\quad {\rm and}\quad
\ket{mn}= 0 \quad {\rm if}~~ m=n$.

\noindent The metric is defined as
$\quad\vev{mn|mn}=\vev{m|m}\vev{n|n}
           =-\vev{mn|nm}
           =\vev{nm|nm},\quad $\\
where  $\vev{m|m}=\vev{\lambda^m|\lambda^m}$ is the
(diagonal) metric element for \CPN.

It follows that the $tt^*$ equations for \Gkn, for any $k$,
 can be derived from the ones for the \CPN\ model.
Defining
\eqn
q_{l_1l_2\dots l_k}&=&\ln g_{\overline{l_1l_2\dots l_k},l_1l_2\dots
l_k} - {{2\sum_i l_i -k(N-1)}\over{2N}}\log|\beta|^2\no\\
&=& q_{l_1}+q_{l_2} +\dots + q_{l_k}.
\enq
The $tt^*$ equations become
\eq\label{kn}
\del_{\bar{z}}\del_zq_{l_1l_2\dots l_k}+
\sum_{i=1}^k\Big\{\exp[q_{l_1l_2\dots l_{i+1}\dots l_k}-q_{l_1l_2\dots l_k}]-
\exp[q_{l_1l_2\dots l_k}-q_{l_1l_2\dots l_{i-1}\dots l_k}]\Big\}
=0
\en
with
\eq
0\leq l_1< l_2<\dots<l_i<\dots <l_k\leq N-1,
\en
and any exponential containing any $q$ with 2
indices the same is ignored.

We now discuss the boundary conditions, which are derived from the UV
and IR limits of those for the \CPN model.
For \CPN,  all the fundamental
one-soliton contributions to the $tt^*$ metric in the IR can be
determined
\eq
q_i=\sum_{r=1}^{N-1} {N \choose r}\sin\Big[{{2\pi r}\over
N}(i+\half)\Big]{1\over \pi}
K_0(m_r\beta) +\dots
\en
for $i=0,\dots ,N-1$, with masses
$m_r=4N|\beta|^{1\over N}\sin \Big({{\pi r}\over N}\Big).$

\noindent Then the boundary conditions in the IR for \Gkn\
are  simply given by
\eq
q_{l_1l_2\dots l_k}=q_{l_1}+q_{l_2}+\dots q_{l_k}
\en
This allows to obtain the soliton spectrum  and soliton numbers
between various
vacua from the canonical basis,\cite{cv2,b1}
the basis of vacua for \Gkn. For  example we find that, for \G24,
solitons fall into multiplets of completely antisymmetric
representations of $SU(4)$, the representations and (the two different)
masses being determined by how far apart the respective six vacua are
(see \cite[x]{b1}).

We are now interested in checking whether the $tt^*$ equations
become solvable in the large $N$ and $k$ limit, in the same way as they were
for the large $N$ \CPN model.\cite{bd}

\noindent Consider first the equations for \Gt\ in the large-$N$
limit. There is an immediate generalization
to \Gkn\ for  any $k$.
Again we can assume that the metric  becomes a continuous function
of  two variables $s_1\equiv i/N$, and $s_2\equiv j/N$.
Redefining
\eq
q_{ij}={1\over N}\ln\vev{\bar{i}|i}\vev{\bar{j}|j}
+2{{(i+j)}\over{N}}\log|\beta|^2,
\en
the $tt^*$ equations become
\eq
{4\over N}\del_{\bar{\beta}}\del_\beta ~q_{ij}+e^{N(q_{ij+1}-q_{ij})}
-e^{N(q_{ij}-q_{ij-1})}+e^{N(q_{i+1j}-q_{ij})}-e^{N(q_{ij}-q_{i-1j})}=0
\en
with $q_{i+N,j}=q_{ij}$ and $q_{i,j+N}=q_{ij}$.
Then, expanding,
the equation reduces to
\eq\label{eq}
4\del_{\bar{\beta}}\del_\beta\,q(s_1,s_2)
={\del\over{\del s_1}}\exp{\Big[{{\del q}\over{\del s_1}}\Big]}
+{\del\over{\del s_2}}\exp{\Big[{{\del q}\over{\del s_2}}\Big]}
\en
with general solution
$q(s_1,s_2)=\ln g(s_1,s_2)= \ln [g(s_1)g(s_2)]=q(s_1)+q(s_2)$,
where $s_2>s_1$, and
 $q(s_i)$ is the solution for the \CPN model
in the large $N$ limit.

The generalization for arbitrary (finite or
infinite)  $k$ is, for  \Gkn\
\eq
\del_{\bar{z}}\del_z\,q(s_1,s_2,\dots,s_k)
=\sum_{i=1}^k{\del\over{\del s_i}}\exp{\Big[{{\del q}\over{\del s_i}}\Big]},
\en
with solution
$q(s_1,s_2,\dots,s_k)=\ln g(s_1,s_2,\dots,s_k)=\ln [g(s_1)g(s_2)\dots
g(s_k)]$,\quad and $s_1<s_2<s_3<\dots <s_k$.

\noindent We note that there is no particular
difficulty in going beyond  $k=1$.

The Grassmannian $\sigma$-models have not been solved
completely as quantum field theories and these results should provide
further insights into them.

\section{The perturbed $A_n$ Minimal models}

The flow of unitary
models towards other massless unitary models were first studied
in perturbation theory.\cite{zam,cl} It was shown that
perturbing the $(m+1)^{\rm th}$ minimal model (for large $m$)
by the least relevant
operator would make it flow in the IR to the $m^{\rm th}$ minimal
model, as the coupling to this operator grows. In this way, the authors
found a flow between successive minimal models.
The extension of these results to the RG flow in $N=1$ discrete series
was discussed in \cite[x]{kms}.
These integrable deformations have also been studied using
the TBA\cite{zam1}.

Here we look at the $tt^*$ formalism for the $N=2$ minimal
models $\,\,W=X^{n+1}/n+1\,\,$ perturbed
by different supersymmetry preserving
operators.\cite{cv1}
By considering various
perturbations $tX^{k+1}$, it is possible to study  non-perturbatively
(for large $n$)
the flow between minimal models  and other minimal models and massive
models.
The integrable models for the minimal models perturbed by
the two  most-relevant supersymmetry preserving operators
can be solved completely, and the
application  to non-integrable perturbations is briefly
discussed.

Consider first the  Landau-Ginsburg potential
corresponding to the $N=2$ $A_n$ series
deformed by the most relevant chiral operator\cite{cv1}
\eq
W(X,t)={{X^{n+1}}\over{n+1}} - t X
\en
where $X$ is the non-trivial chiral field of lowest dimension.

The ring is identical to the one for the \CPN model,
and the $tt^*$ equations are thus of the same affine Toda form
\eq
     \partial_z\partial_{z^*}q_i +
          e^{(q_{i+1}-q_i)} -
        e^{(q_i-q_{i-1})}=0
\en
with
\eqn
q_i&=&\log\langle\bar i|i\rangle -{{2i-n+1}\over{2n}}\log|t|^2\no\\
 z &=& {n\over {n+1}}t^{{(n+1)}\over n}\no\\
 q_i&\equiv & q_{i+n},\quad 0=q_i + q_{n-i-1}.
\enq
We will again consider the large $n$ limit of these equations. We note now
the difference with \CPN.
For the \CPN\ model, the large $N$ effective action still has $N$
fields in the
theory and the $q_i$ were  rescaled by $1/N$. Here we do not
need this rescaling.
To get a smooth large $n$ limit we redefine the $q_i$'s in
terms of a continuous variable $s\equiv {i\over n}$,
\eqn
q_i&\rightarrow& q(s,|t|) =\log g(s,|t|)-(2s-1)\log|t|,\no\\
z&\rightarrow& nt
\enq
The  $tt^*$ equations become
$\qquad\quad\partial_{t}\partial_{{t}^*}{1\over n} F
+{{\partial^2}\over{\partial s^2}}e^{{1\over n}F}=0$, \quad

\noindent with $F= {{\del q}\over{\del s}}.$
The exponential is of order ${1\over n}$, and we get
\eq\label{lap}
\partial_{t}\partial_{{t}^*}F + {{\partial^2}\over{\partial s^2}}
F =0,
\en
 the ordinary 3D Laplace equation.
The  general solution for the metric is
\eq
q(s,t)=-\sum_{p\geq 1}{{c_p\sin 2\pi p s}\over{\pi p}}~K_0(4\pi p|t|)
+ (as + b)\log |t|
\en
where $c_p$ is the soliton multiplicity and is 1 (see \cite[x]{fi1}).\\
\noindent In the limit where  $t\rightarrow 0$
\eq
q(t,s)\rightarrow -\sum_{p\geq 1}{{\sin 2\pi p s}\over{\pi p}}~
       [-\ln(2\pi p|t|)-\gamma] + (as+b)\log |t|
\en
Then, we can use the expression\cite{mangulis}
\eq
\sum_{p\geq 1}{{\ln p}\over{p}}~\sin 2\pi p s
= \pi\ln\Gamma (s) + {\pi\over 2}\ln(\sin\pi s) -
\half \pi(1-2s)(\gamma + \ln 2) -\pi(1-s)\ln\pi
\en
for  $0<s<1,\quad$
and
\eq
\sum_{p\geq 1}{{\sin 2\pi p s}\over{p}}=\pi(k+\half -s)\qquad {\rm for\quad}
2k\pi<2\pi s<(2k+1)\pi.\en
We have  $\qquad q(t,s)= \ln{\Gamma^2(s)\sin \pi s}-\ln\pi -(2s-1)\log |t|
+ (as+b)\log|t|\qquad$
\eq\label{limit}
{\rm and}\qquad  g(s,t)={1\over \pi}\Gamma^2(s)\sin\pi s
\en
for $a=b=0$.
\noindent This is consistent with the known result\cite{cv1} for the
ground state metric at $t =0$
\eq\label{min}
 g_{\bar{i}i}(t\rightarrow 0)=
{{\Gamma ({{i+1}\over {n+1}})}\over{\Gamma (1-{{i+1}\over {n+1}})}},
\en
which for large $n$  becomes  (\ref{limit}).
In the IR (large $t$), we have
\eq\label{as}
q\sim -\sum_p \sin 2\pi p s{{e^{-4\pi p|t| }}\over{\sqrt{8\pi^2p^3|t|}}}
+..
\en
We can now compare this with the results in \cite[x,y]{cv1,b1}
obtained from semi-classical considerations
\eq
q_i\sim -\sin\left[{{2\pi}\over n}(i+\half)\right]{{e^{-4n|t|\sin{\pi\over
n}}}\over{\sqrt{8\pi n|t|\sin{\pi\over n}}}}.
\en
In the large $n$ limit this is
$\quad-\sin(2\pi s){{e^{-4\pi |t|}}\over{\sqrt{8\pi^2|t|}}},\quad$
the first term in (\ref{as}).

In order to derive the `new index', we  write the metric in terms of
 a dimensionless quantity, $z=M\beta$, where $M=4\lambda$ is
the mass of the fundamental soliton\cite{fi1}  and $\beta$
is the perimeter of the cylinder
(set to 1 up to now). This is done by
rescaling the superfields $X\rightarrow t^{1\over n}X$;
then
$W\rightarrow \lambda W$ with $\lambda=t^{{1+n}\over n}\,\,$ and
$\,\,q_i(z) =\log\langle\bar i|i\rangle $
with $\det [g]=1.$ The action of the renormalization group can  be
seen as this rescaling of the superpotential by $\lambda$. Then
 $\lambda\rightarrow 0$ corresponds to the UV limit and
 $\lambda\rightarrow \infty$ to the IR.

In the large $n$ limit, $\lambda \rightarrow t\quad$  and
$\quad q(s,z)=\log g(s,z),\quad$
and the ground-state metric obeys
\eq
  16\del_{z}\del_{{z}^*} {{\del q}\over{\del s}}
+{{\partial^2}\over{\partial s^2}}{{\del q}\over{\del s}}=0,
\en
\eq
{\rm and}\qquad\qquad q(s,z)=-2\sum_{p\geq 1}{{\sin 2\pi p s}
\over{\pi p}} ~K_0(\pi p |z|)+...
\en
The dependence of the metric on the coupling, as
$z\rightarrow 0$,
$g\rightarrow {1\over \pi}\Gamma^2(s)\sin\pi s ~
 |z|^{1-2s}$,
should be  consistent with the result\cite{cv1}
that in the basis of ground-state vacua of definite
charge the metric behaves as
$\quad g_{\bar{i}i}\sim |z|^{-2q_i},\quad$
where $q_i$ is the charge of the $i$-th Ramond vacuum.
Thus we find  the Ramond charge to be, in the large $n$ limit,
$\quad q^R_s=s-\half$.
The Neuveu-Schwarz $U(1)$ charges for the $N=2$ minimal models
are  $q_k={{k}\over{(n+1)}}$
for $k=1,..., n$, which  become $q_s=s$, and
since the Ramond charges are $q^R_k=q_k-\hat{c}/2$,
this tells us that $\hat{c}=c/3$ or $c=3$.

We now look at  the index $Q$. In the conformal limit, the
eigenvalues $Q_{ab}$ of the index
should be the same as the $U(1)$ charges for the
superfields $|X^k\rangle$ in the Ramond sector for the $N=2$ minimal
models.

\noindent The index $Q(s,z)$ is simply\cite{cfiv}
\eq
Q(s,z)=-{z\over 2} g^{-1}\del_z g=-\sum_{p\geq 1}{z\sin 2\pi p s}
{}~K_1(\pi p|z|)
\en
We see that in the IR, we get a series of terms in $e^{-pM\beta}$ each
of which is the contribution of a single massive particle saturating
the Bogolmonyi bound $m=2p|\Delta W|$. There are no corrections from
multi-solitons states, which seem to suggest that the particles remain
free. Actually this is not too surprising since the coupling and
non-trivial scattering for large $n$ is  of order $O(1/n)$.

In the UV limit $z\rightarrow 0$,
\eq
Q(s,0)\rightarrow  -\sum {{\sin{2\pi p s}}\over{\pi p}}\rightarrow s -\half
\en
for all $s$, which is the desired result.
This is also
consistent with the well-known result that
the ground-state energy for the minimal models is given by\cite{cardy}
\eq
E(\beta)=-{\pi\over{6\beta}}\left[{{3(n-1)}\over{(n+1)}}\right]
={{2\pi}\over{\beta}}(-{c\over{12}})
\en
which for large $n$ implies $c=3$.

As mentionned earlier, at first order the particles seem free. However
we can generate corrections of order
$1\over n$.
Call $F^{(0)}$ the first order solution of (\ref{lap})
\eq
16\del_z\del_{z^*} F^{(0)} + {{\del^2}\over{\del s^2}} F^{(0)} =0
\en
with
\eq
F^{(0)}=-4\sum \cos 2\pi p s~K_0(\pi p |z|)
\en
The next order correction in the exponential expansion will be a
solution of
\eq
{4\over{z}}\del_z z\del_z F^{(1)}+\half {{\del^2}\over{\del
s^2}}[F^{(0)}]^2=0
\en
Integrating, using formulas found in \cite[x]{grad},
\eq
F^{(1)}=\sum_{p,p'}{{8\sin 2\pi p s ~\cos 2\pi
p's}\over{(p^2-p'^2)}}\left[pp'K_0(\pi p|z|)K_1(\pi
p'|z|)+p^2K_0(\pi p'|z|)K_1(\pi p |z|)\right]
\en
Integrating over $s$, one finds
\eqn
q^{(1)}&=&\sum_{p,p'}{-4\over{(p^2-p'^2)}}\left[{{\cos 2\pi
s(p+p')}\over{(p+p')}}
+{{\cos 2\pi s(p-p')}\over{(p-p')}}\right].\\
  &&.\left[pp'K_0(\pi p|z|)K_1(\pi
p'|z|)+p^2K_0(\pi p'|z|)K_1(\pi p |z|)\right]
\enq
which gives  the two-particle interactions.\cite{cfiv}
In such a way, it is possible to generate successive corrections to
the index of  order $1\over n$ and determine the $S$-matrix.

So far we have looked at the $N=2$ Minimal models perturbed by the
most relevant chiral field.
Consider now a general perturbation of the form
\eq
W={{X^{n+1}}\over{n+1}}-t{{X^{k+1}}\over{k+1}}
\en
The critical points are $X^n=tX^k$
and  $X=0$ is degenerate of order $k$. As we move away from
the conformal theory at $t=0$, the critical points spread out, and in the
IR limit, for large $t$, we have  a set of isolated critical
points.
The flow generated by the
original model has taken us to another massless model in the IR limit
around $X=0$ (with $W=X^{k+1}$). Around the other critical points the
models are massive in the IR limit and  defined by
$W={{X^{n-k+1}}\over{n-k+1}}-tX$.

The $tt^*$ equations can be studied for any
direction of perturbation, however only very
special directions lead to integrable quantum field theories and these
cases produce $tt^*$ equations of the affine Toda type.\cite{cv1}
These equations  always lead to massive theories in the
IR. For $k>1$, the models are not integrable and the $tt^*$ equations
become more complicated, also, solitons between vacua do not always
saturate the Bogolmonyi bound.

For a general perturbation $X^{k+1}$, the coupling matrix $C_t$
which characterizes the chiral ring by
multiplication by the operator
$-\del W/\del t$ has components
\eqn
C_{ij} &=& {-{1\over {k+1}}}\delta_{j-i,k+1}~~~~~~~~~~~{\rm for}~~
     i=1,2,\dots ,n-1-k\\
     &=&{-{t\over{k+1}}}~\delta_{j-i, 2k+1-n}~~~~~{\rm for}~~i=n-k,\dots,n.
\enq
The model is invariant under the  symmetry
$\quad X\rightarrow e^{{2\pi i}\over{n-k}}X,\quad $
which leaves the chiral elements $\{1,X^{n-k}\}$ invariant.
Under this symmetry, the elements
$\{X^l,X^{n-k+l}\}$ also transform in the same way.
Thus the metric $g_{\bar{h},k}=\langle
h|\bar{k}\rangle$ is diagonal except for elements $\langle{l}|{n-k+l}\rangle$.

\noindent The topological metric is the 2-point correlation function
\eq
\eta_{ij}=\langle\phi_i\phi_j\rangle_{top}=\langle i|j\rangle
={\rm Res_W}[\phi_i\phi_j]
\en
with
\eq
{\rm Res_W}[\phi]={1\over{(2\pi
i)^n}}\int_{\Gamma}{{\phi (X)dX^1\wedge\dots\wedge dX^n}
\over{\del_1W\del_2W\dots\del_nW}}
\en
And we have
\eq
\eta_{ij}=\delta_{i+j,n-1}+t~\delta_{i+j,2n-1-k}
+t^2~\delta_{i+j,3n-1-2k}+\dots
\en
The first perturbation, $tX^2/2$, is integrable\cite{flmw}
and the finite $n$ case has been discussed in \cite[x]{cv1}.
The topological metric  $\eta_{ij}=\delta_{i+j,n-1}+t~\delta_{i+j,2n-2}$.
In the large $n$ limit, the model can be solved completely,
the $tt^*$ equations reducing again to the Laplace equations (with different
coefficients). We will not discuss the solution in detail but just
mention that one needs to define two sets of equations since the
equations couple metrics of odd $i$ together and metrics of even $i$
together.

Going beyond $k=1$ becomes more complicated. It is not yet clear
whether the equations admit a nice solution in the large $n$ limit.
Consider the perturbation $W={{X^{n+1}}\over{n+1}}-t{{X^3}\over 3}$.
The boundary conditions are known. In the UV, we  have (\ref{limit}). In
the IR, we should expect a combination of (\ref{as}) for the massive model
and (\ref{min}) (with $i=0,1$) for the massless model.

\noindent We can choose to
work in the `flat coordinates' basis\cite{blok,verlinde}
\eq
\{1,X,X^2,\dots,X^{n-2}-{t\over 3},X^{n-1}-{2\over 3}Xt\}
\en
The elements $\{1,X^{n-2}\}$ and
$\{X,X^{n-1}\}$ change in the same way under
$X\rightarrow \exp ({{2\pi i}\over {n-2}})X$
and the reality constraints (resulting from the relation
$\eta^{-1}g(\eta^{-1}g)^*=1$
between $g$ and $\eta$) imply
\eqn
g_{0\overline {n-2}}g_{1\bar{1}}+g_{0\bar{0}}g_{1\overline{n-1}}&=&0;\quad
g_{\overline {n-2}0}g_{n-1\overline{n-1}}+g_{n-2\overline{n-2}}
g_{n-2\bar{0}}=0\\
g_{n-2\overline {n-2}}g_{1\bar{1}}+g_{n-2\bar{0}}g_{1\overline{n-1}}&=&1;\quad
g_{n-2\overline {n-2}}g_{1\bar{1}}+g_{n-2\bar{0}}g_{1\overline{n-1}}=1
\enq
The $tt^*$ equations then become more complicated.
For example the first equation is
\eq
\del_{\bar{t}}\left(g_{0\bar{0}}\del_tg_{n-1\overline{n-1}}+
g_{0\overline{n-2}}\del_tg_{n-1\bar{1}}\right)={1\over 9}
\left[g_{3\bar{3}}g_{n-1\overline{n-1}}+{2\over 3}\bar{t}g_{3\bar{3}}
g_{n-1\bar{1}}\right].
\en
It is not obvious at this point how or if the equations decouple
in these cases. We hope to
study this question further.

\section{Conclusion}
We have seen here that the $tt^*$ formalism is a powerful tool
in unravelling  non-trivial dynamical
information about a full quantum field theory from its topological
data.
The formalism is
completely solvable for various models
with a large number of
superfields.
One could hope  that these methods
will also have applications in $D>2$.

\section{Acknowledgements}
The work on the \CPN model was done in collaboration with
M. Douglas.
I would like to thank J. Cardy, M. Douglas, P. Fendley,
 K. Intriligator and R. Sorkin for useful and enjoyable discussions.
I would also like to thank the organisers  for their
efforts in making the conference a success.

\section{References}

\end{document}